\documentstyle[aps,epsfig,multicol,eqsecnum]{revtex}   

\newcommand{\s}[1]{$^{(#1)}$}
\newcommand{\zx}{\zeta(x_1)}
\newcommand{\tm}{\theta_m}
\newcommand{\ew}{\epsilon(\omega)}
\newcommand{\ts}{\theta_s}
\renewcommand{\to}{\theta_0}
\newcommand{\p}{\partial}

\newcommand{\ci}{^{\circ}}
\newcommand{\bqe}{\begin{eqnarray}}
\newcommand{\eqe}{\end{eqnarray}}
\newcommand{\nn}{\nonumber}
\newcommand{\fr}{\frac{1}{2}}

\newcommand{\w}{\omega }
\newcommand{\prps}{\left\la\frac{\p R}{\p\ts}\right\ra}

\newcommand{\la}{\langle}
\newcommand{\ra}{\rangle}

\newcounter{tr}

\newcommand{\sctr}[1]{\setcounter{tr}{#1}}
\newcommand{\sceq}{\addtocounter{equation}{-1}}

\begin{document}
\title{
 Random Surfaces that Suppress Single
Scattering }
\author{A. A. Maradudin and I. Simonsen}
\address{
Department of Physics and Astronomy
and Institute for Surface and Interface Science,\\
University of California,
Irvine, CA 92697, U.S.A.}
\author{T. A. Leskova}
\address{
Institute of Spectroscopy,
Russian Academy of Sciences,
Troitsk 142092, Russia}
\author{E. R. M\'endez}
\address{
Divisi\'on de F\'{\i}sica Aplicada,
Centro de Investigaci\'on Cient\'{\i}fica y de Educaci\'on
Superior de Ensenada,\\
Apartado Postal 2732,
Ensenada, Baja California 22800, M\'exico}
\maketitle
\begin{abstract}
We present a method for generating numerically a one-dimensional random
surface, defined by the equation $x_3 = \zx$, that suppresses
single-scattering processes in the scattering of light from it within a
specified range of scattering angles.  Rigorous numerical calculations of
the scattering of light from surfaces generated by this approach show that
the single-scattering contribution to the mean scattered intensity is
indeed suppressed within that range of  angles.
\end{abstract}
\newpage

In theoretical and experimental studies of multiple-scattering effects in
the scattering of light from randomly rough surfaces it is often desirable
to be able to suppress the contribution to the mean scattered intensity
from single-scattering processes:  effects such as enhanced backscattering
or the presence of satellite peaks become more readily observable in the
absence of the background provided by single-scattering processes.

In theoretical studies it is possible to separate the contribution from
single-scattering processes to the mean intensity of the light scattered
incoherently from the contribution from multiple-scattering processes.
This separation of single- and multiple-scattering contributions is
particularly simple in the context of small-amplitude perturbation
theory\s{1}, and not much more difficult in a computer simulation
approach\s{2}.  However, it is not so easy to achieve experimentally.  In
the case of the scattering of light  from  two-dimensional random surfaces,
the in-plane, cross-polarized scattering of p-polarized light suppresses
the single-scattering contribution to the mean intensity of the light
scattered incoherently.  In the case of the scattering of light incident
normally on  a weakly rough one-dimensional random metal surface, the use
of a surface whose roughness is characterized by a power spectrum $g(|k|)$
that vanishes identically for $|k| < k_{min} \leq \w /c$ eliminates the
contribution from single-scattering processes to the mean intensity of the
incoherent component of the scattered light for scattering angles smaller
in magnitude than $\sin^{-1}(ck_{min}/k\w )$ \s{3}.  However, such surfaces
are difficult to fabricate.

In this paper we explore a different approach to the design of random
surfaces that suppress the single-scattering contribution to the incoherent
component of the light scattered from them, that is not restricted to
weakly rough surfaces, and that appear to be easier to fabricate than
surfaces characterized by a West-O'Donnell power spectrum.

To motivate this approach, let us consider the scattering of an s-polarized
plane wave of frequency $\w$ from a one-dimensional, perfectly conducting,
random surface, when the plane of incidence is perpendicular to the
generators of the surface.  We recall that if the inhomogeneous Fredholm
equation for the normal derivative of the single nonzero component of the
electric field in the vacuum, evaluated on the surface, is solved by
iteration, the inhomogeneous term yields the Kirchhoff approximation to the
mean scattered intensity, a single-scattering approximation, the first
iterate yields the pure double-scattering contribution, and so on\s{2}.
Consequently, if a surface can be designed with the property that the
Kirchhoff approximation to the mean intensity of the light scattered from
it vanishes for the scattering angle $\ts$ in the interval $(-\tm ,\tm )$,
all the scattered intensity within this range of scattering angles will be
due to multiple-scattering processes.  Consequently, our aim is to design a
one-dimensional, perfectly conducting random surface for which the
Kirchhoff approximation to the mean differential reflection coefficient
vanishes identically for $\ts$ in the interval $(-\tm ,\tm )$.  The
analysis required is simplified significantly by working in the geometrical
optics limit of the Kirchhoff approximation.  However, the results obtained
still display the behavior sought.

Thus, we consider a one-dimensional, randomly rough, perfectly conducting
surface defined by the equation $x_3 = \zx$, that is illuminated by an
s-polarized plane wave of frequency $\w$.  The surface profile function
$\zx$ is written in the form \s{4}
\bqe
\zx = \sum^{\infty}_{\ell = - \infty} c_{\ell}s(x_1-\ell 2b) ,
\eqe
where the $\{ c_{\ell}\}$ are independent, positive, random deviates, $b$
is a characteristic length, and the function $\zx$ is defined by\s{4}
\bqe
s(x_1) &=& \left\{
\begin{array}{ll}
0, &\quad x_1 < - (m+1)b,\\
 -(m+1)bh-hx_1, &\quad - (m+1)b < x_1 < -mb,\\
 - b, &\quad -mb < x_1 < mb,\\
-(m+1)bh+hx_1, &\quad mb < x_1 < (m+1)b,\\
 0, &\quad (m+1)b < x_1 ,
\end{array}
\right.
\eqe
where $m$ is a positive integer.  Due to the positivity of the coefficient
$c_{\ell}$, its probability density function (pdf) $f(\gamma ) = \la \delta
(\gamma - c_{\ell})\ra$ is nonzero only for $\gamma > 0$.

It has been shown that for the random surfaces defined by Eqs. (1) and (2)
the mean differential reflection coefficient in the geometrical optics
limit of the Kirchhoff approximation is given by\s{4}
\bqe
\prps = \frac{1}{2h} \frac{[1+\cos (\to +\ts
)]^2}{\cos\to (\cos\to + \cos\ts )^3}\left[ f\left( \frac{\sin\to -\sin\ts
)}{h(\cos\to + \cos\ts )}\right) +
f\left( \frac{\sin\ts - \sin\to}{h (\cos \to +
\cos\ts )}\right)\right] ,
\eqe
where $\to$ and $\ts$ are the angles of incidence and scattering, measured
counterclockwise and clockwise from the normal to the mean scattering
surface, respectively.  This result shows that $\la \p R/\p\ts\ra$ is given
in  terms of the pdf of the coefficient $c_{\ell}$ and is independent of
the wavelength of the incident light.  It simplifies greatly in the case of
normal incidence $(\to = 0\ci )$,
\bqe
\prps = \left( 1 + \tan^2\frac{\ts}{2}\right)
\frac{f(-\frac{1}{h}\tan\frac{\ts}{2})+f(\frac{1}{h}\tan\frac{\ts}{2})}{4h}
,\nn\\
\eqe
and we will restrict ourselves to this case in what follows.  From Eq. (4)
we find that if we wish $\la \p R/\p\ts\ra$ to have the form, say,
\bqe
\prps = \left\{\begin{array}{ll}
0, &\quad 0 < |\ts | < \tm,\\
\frac{\cos\ts}{2(1-\sin\tm )}, &\quad \tm < |\ts | < \pi /2 ,
\end{array}
\right.
\eqe
we must choose for $f(\gamma )$
\bqe
f(\gamma ) =\left\{\begin{array}{ll} 
0, &\quad  0 < \gamma < \gamma_m,\\
2h \frac{1+h^2\gamma^2_m}{(1-h\gamma_m)^2}
\frac{1-h^2\gamma^2}{(1+h^2\gamma^2)^2},&\quad \gamma_m < \gamma <
\frac{1}{h} ,
\end{array}
\right.
\eqe
where $\gamma_m = [\tan (\tm /2)]/h$.  From this form for $f(\gamma )$ a
long sequence of $\{ c_{\ell}\}$ can be generated, e.g. by the rejection
method\s{5}, and the surface profile function generated by the use of Eqs.
(1) and (2).

The surface profile functions $\zx$ generated in this way are not zero-mean
Gaussian random processes, and are not stationary.  Indeed, the mean square
height of the surface, $\delta^2 = \la \zeta^2(x_1)\ra - \la \zx \ra^2$, is
a  periodic function of $x_1$ with a period $2b$, and for $m = 1$ is given
by $\delta^2 = [\la c^2\ra - \la c\ra^2]h^2b^2 [1+(x_1/b)^2]$ for $-b\leq
x_1 \leq b$.  The average of this function over a period, $\delta^2_{av} =
[\la c^2\ra - \la c\ra^2]4h^2b^2/3$, can be used to estimate the rms height
of the surface.  Similarly, the mean square slope of the surface is given
by $s^2 = \la (\zeta '(x_1))^2\ra - \la \zeta '(x_1)\ra^2$ $ = [\la c^2 \ra
-\la c\ra^2]h^2$, from which the rms slope can be determined.  The averages
$\la c\ra$ and $\la c^2\ra$ appearing in these expressions, the first two
moments of $f(\gamma )$, are given by
\sctr{1}
\bqe
\la c\ra = \frac{1}{h} \frac{1+h^2\gamma^2_m}{(1-h\gamma_m)^2} \left\{ \cos\tm
+ 2\ln \frac{\cos \left(\frac{\pi}{ 4}\right)}{\cos \left(\frac{\tm }{2}\right)}\right\}\\ \sctr{2}\sceq
\la c^2\ra = \frac{2}{h^2} \frac{1+h^2\gamma^2_m}{(1-h\gamma_m)^2}
 \left\{
\frac{\pi}{2} - \tm +\tan \left(\frac{\tm}{2}\right) 
\left[1+\cos^2\left(\frac{\tm }{2}\right)\right] - \frac{3}{2}\right\}.
\eqe

An example of a surface generated in this way is presented in Fig. 1.  The
pdf $f(\gamma )$ used in its generation is the one defined by Eq. (6), with
$\tm = 40.1  \ci$.  The parameters entering the definition of the function
$s(x_1)$ are $b = 3 \lambda$, $m = 3 $, and $h = 0.2.$ For these values of the parameters we find that $\delta_{av}=0.9\lambda$, and $s=0.57$, so that
the surface is moderately rough.  It was
sampled at the points $x_p = [(p+\fr )b]/N$, where $p = 0, \pm 1, \pm
2,\ldots$ and $N = 100$, and  is seen to consist of a succession of
triangular peaks and valleys.

To show that this random surface suppresses single-scattering processes for
$|\ts | < 40.1 \ci$, we have plotted in Fig. 2 the contribution to the mean
differential reflection coefficient from the incoherent component of the
scattered light, $\la \p R/\p\ts\ra_{incoh}$, for scattering from this
surface, calculated by a computer simulation approach in the Kirchhoff
approximation\s{2}, with and without invoking the geometrical optics limit
of the latter.  The results for a total of 2000 realizations of the surface
were used in carrying out the ensemble average required for obtaining the
mean differential reflection coefficient.  It is seen that in the
geometrical optics limit of the Kirchhoff approximation $\la \p R/\p
\ts\ra_{incoh}$ vanishes for $|\ts | < 40.1 \ci$.  In the Kirchhoff
approximation $\la \p R/\p\ts\ra_{incoh}$ is not identically zero in this
region of scattering angles,  but is quite small.  The difference between
these two results shows how well the geometrical optics limit of the
Kirchhoff approximation reproduces the result of the Kirchhoff
approximation itself.   In this figure we have also plotted the total
contribution to the mean differential reflection coefficient from the
incoherent component of the scattered light, including all
multiple-scattering contributions.  This result for $\la \p
R/\p\ts\ra_{incoh}$ was calculated exactly by a computer simulation
approach\s{2}, in which the results for 2000  realizations of the surface
were averaged.  We see from this figure that there is now a low background
for $|\ts | < 40.1 \ci$, due to multiple-scattering, on which is
superimposed an enhanced backscattering peak in the retroreflection
direction $(\ts = 0\ci )$, whose height is nearly twice that of the
background at its position.  The latter is the result expected when the
contribution from single-scattering processes has been subtracted\s{6}.

Although the theory underlying the approach to generating random surfaces
that suppress single scattering presented here was based on the assumption
that the scattering surface is perfectly conducting, the resulting approach
also works very well for finitely conducting surfaces.  In Fig. 3  we have
plotted a rigorous computer simulation result for $\la \p
R/\p\ts\ra_{incoh}$ in the case that s-polarized light of wavelength
$\lambda = 612.7$ nm is incident normally on a one dimensional random
silver surface $(\ew = - 17.2+i0.498 )$.  The surface is characterized by
the parameters $b = 3 \lambda$, $m = 3, h = 0.2$, and $\tm = 40.1\ci$.
Results for 2000 realizations of the surface were averaged in obtaining
this figure.  The strong suppression of $\la\p R/\p\ts\ra_{incoh}$ in the
interval $|\ts | < 40.1 \ci$ is clearly seen, and an enhanced
backscattering peak at $\ts = 0\ci$ rises to about twice the height of the
background at its position.

In this letter we have presented a method for generating numerically a
one-dimensional random surface profile function $\zx$ that has the property
that it suppresses the single scattering of s-polarized light from it, a
property that in the case of a perfectly conducting surface is independent
of the wavelength of the incident light.  The extension of the present
approach to the generation of one-dimensional random surfaces that suppress
the single scattering of p-polarized light is straightforward.  The method
described is not restricted to the generation of weakly rough surfaces.
Surfaces defined by Eqs. (1) and (2), with a different form of the pdf
$f(\gamma )$, have been fabricated successfully in the laboratory\s{4}, and
their fabrication appears to be simpler than that of surfaces characterized
by a West-O'Donnell\s{3} power spectrum.  The approach to the design of
two-dimensional random Dirichlet surfaces that act as band-limited uniform
diffusers developed recently\s{7}, can be used to design two-dimensional
random Dirichlet surfaces that suppress single-scattering processes.  These
and other applications of the approach described here will be presented
elsewhere.
\vskip.1in
The work of A.A.M. and T. A. L. was supported by Army Research Office Grant
DAAG 55-98-C-0034. I.S. would like to thank the Research Council 
of Norway (Contract No. 32690/213) and Norsk Hydro ASA for financial 
support. The work of E.R.M. was supported by CONACYT Grant 3804P--A.
This work has also received support from the Research Council 
of Norway (Program for Supercomputing) through a grant of computing time.

\newpage

\begin{figure} 
    \caption{A one-dimensional random surface profile function $\zx$
        obtained from Eq. (1) by the use of the pdf given by Eq. (6)
        together with a function $s(x_1)$ defined by Eq. (2) with $b =
        3 \lambda , m = 3 , h = 0.2$, and $\tm = 40.1 \ci.$}
\end{figure}

\begin{figure} 
    \caption{$\la\p R/\p\ts\ra_{incoh}$ calculated by a computer
        simulation approachfor the random surface displayed in Fig. 1,
        when s-polarized light of wavelength $\lambda$ is incident
        normally on it.  $(\cdots \cdots)$ The geometrical optics
        limit of the Kirchhoff approximation; $( - - - - - )$ the
        Kirchhoff approximation; ($\frac{\mbox{\hspace*{1.5cm}}}{}$) the
        result with all multiple-scattering contributions included.}
\end{figure}

\begin{figure} 
    \caption{$\la \p R/\p\ts\ra_{incoh}$ calculated by a computer
        simulation approach for the case that s-polarized light of
        wavelength $\lambda = 612.7$ nm is incident normally on a
        one-dimensional random silver surface. The arrows indicate the
        positions of the angles $\pm\tm$.}
\end{figure}

\setcounter{figure}{1}
\newcommand{\mycaption}[2]{\begin{center}{\bf Figure \thefigure}\\{#1}\\{\em #2}\end{center}\addtocounter{figure}{1}}
\newcommand{\myauthor}{A. A. Maradudin, I. Simonsen, T. Leskova, and
    E. R. M\'endez}
\newcommand{\mytitle}{Random Surfaces that Suppress Single Scattering }


\newpage

\begin{figure}
    \begin{center}
            \epsfig{file=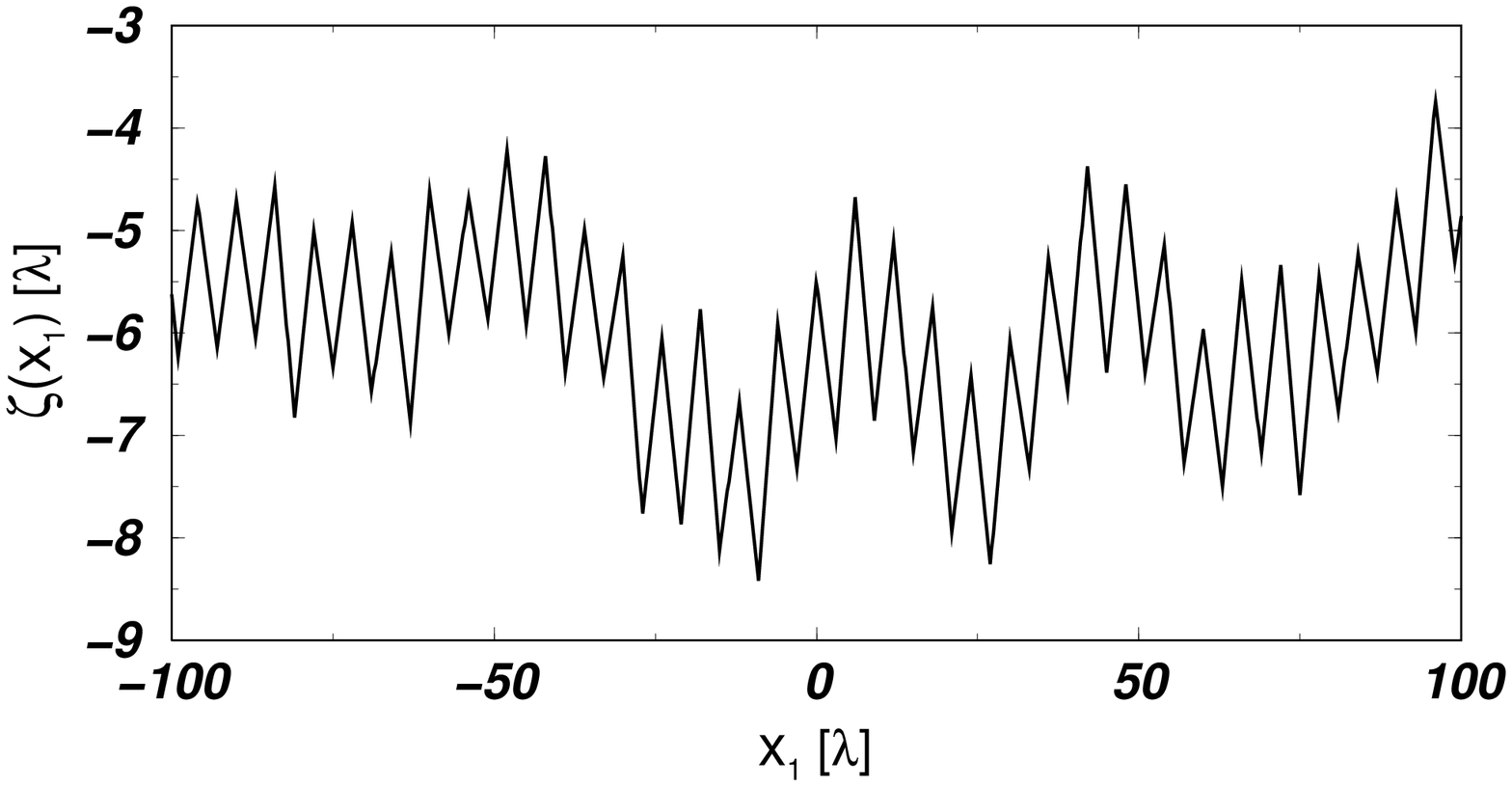,width=8.5cm,height=12.5cm,angle=-90} 
    \end{center}
    \mycaption{\myauthor}{\mytitle}
\end{figure}

\begin{figure}
    \begin{center}
            \epsfig{file=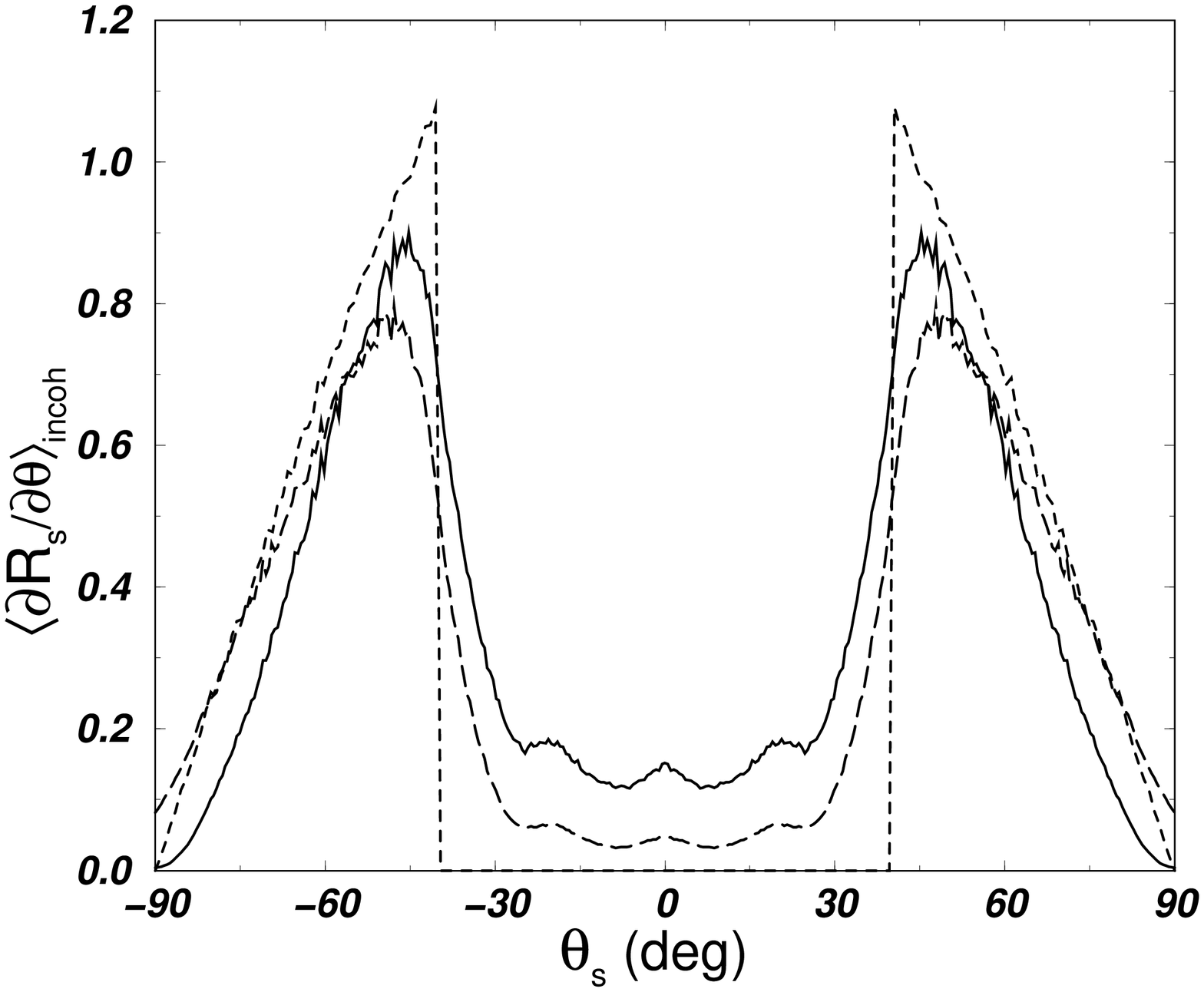,width=9.5cm,height=9.5cm,angle=-90} 
    \end{center}
    \mycaption{\myauthor}{\mytitle}
\end{figure}

\begin{figure}
    \begin{center}
            \epsfig{file=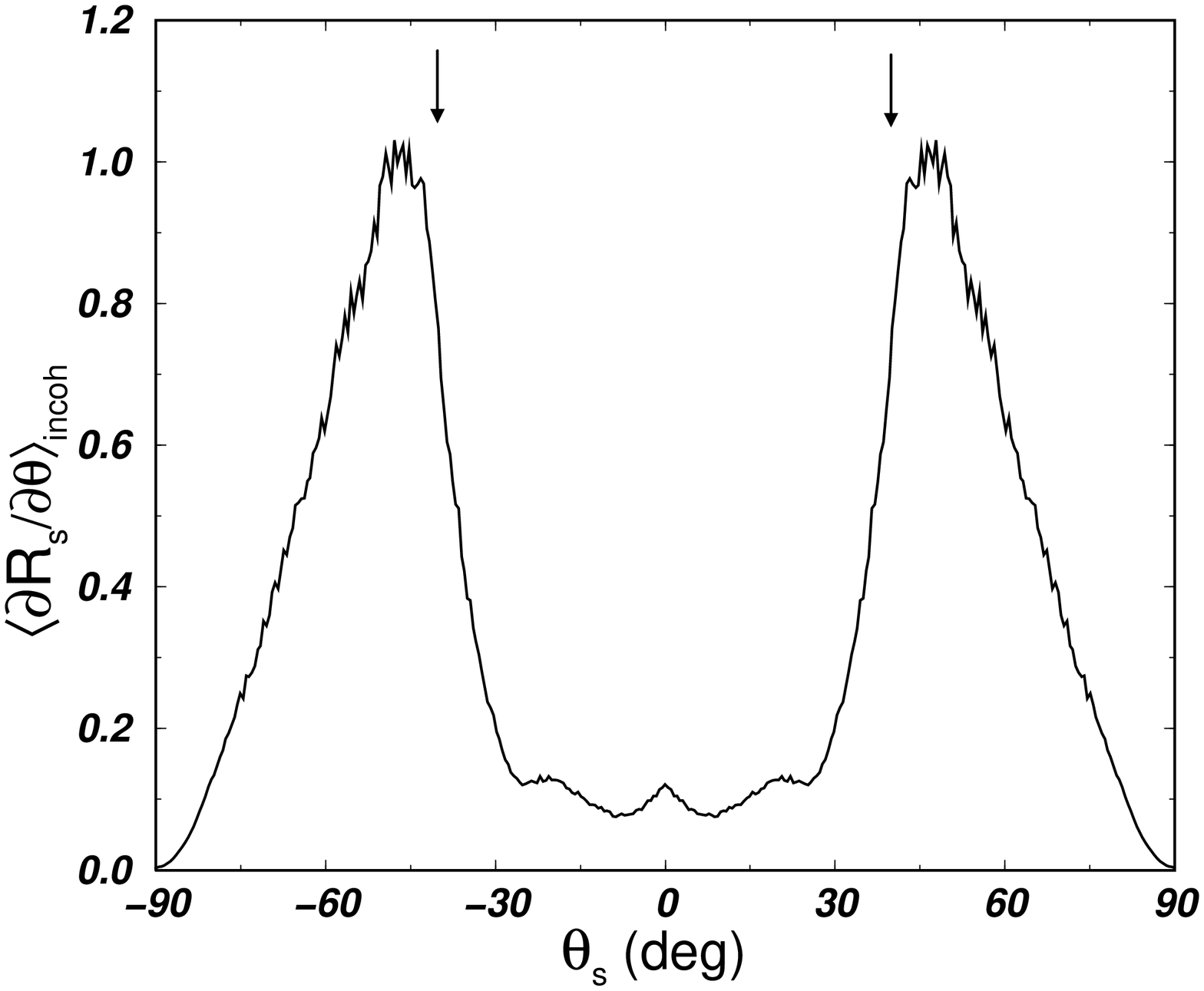,width=9.5cm,height=9.5cm,angle=-90} 
    \end{center}
    \mycaption{\myauthor}{\mytitle}
\end{figure}

\end{document}